# Space and space-time topologies in a type-II hyperbolic lattice

Jingming Chen[1,#], Zebin Zhu[1,#], Minqi Cheng[1,#], Linyun Yang[2], Yuxin Zhong[1], Zhen Gao[1,*]

[1]State Key Laboratory of Optical Fiber and Cable Manufacture Technology, Department of Electronic and Electrical Engineering, Guangdong Key Laboratory of Integrated Optoelectronics Intellisense, Southern University of Science and Technology, Shenzhen 518055, China.
[2]College of Aerospace Engineering, Chongqing University, Chongqing, 400030, China.

#These authors contribute equally
*Corresponding author. Email: gaoz@sustech.edu.cn (Z.G.)

**Recent breakthroughs in hyperbolic lattices have expanded the study of topological phases of matter from Euclidean to non-Euclidean spaces. However, prior work has mostly focused on spatial topological states at the single outer edge of type-I hyperbolic lattices. The dynamic transfer of hyperbolic topological states across multiple edges, as well as the emergence of spatiotemporal topological phenomena, remains largely unexplored. Here, we establish both spatial and spatiotemporal topologies in a newly discovered type-II hyperbolic lattice possessing outer and inner edges. Using electric circuits, we experimentally realize a type-II hyperbolic Chern insulator and directly observe degenerate chiral edge states of opposite chirality at its outer and inner edges. Furthermore, by coupling these counter-propagating chiral edge states, we demonstrate an anti-time-parity phase transition, enabling dynamic transfer between them in arbitrary proportions. Finally, we propose a novel paradigm for constructing a (2+1)-dimensional hyperbolic space-time crystal, which hosts an intertwined topology of spatial Chern and temporal winding numbers, resulting in a unique space-time topological string state. Our work expands the frontier of hyperbolic topological physics, paving the way for the spatiotemporal dynamic manipulation of hyperbolic topological states.**

**Introduction**

The discovery of hyperbolic lattice, a discretized regularization of non-Euclidean space with constant negative curvature, in circuit quantum electrodynamics [1] and electric circuits [2] has opened new avenues to extend topological physics from the Euclidean to non-Euclidean spaces [3-14]. However, to date, previous studies on hyperbolic topological physics have been predominantly restricted to conventional type-I hyperbolic lattices and exclusively focused on the spatial topological states associated with a single outer edge [3-14]. This limitation leaves topological phenomena involving multi-edge spatial interactions and dynamic transitions, such as Laughlin and Thouless pumping [15-18], Landau-Zener transition [19-23], and non-Hermitian phase transition [23-25], largely unexplored. These dynamic topological effects are pivotal for hyperbolic topological physics, as they not only enable dynamic manipulation of hyperbolic topological states but also serve as powerful methods to characterize hyperbolic topological invariants and explore higher-dimensional hyperbolic topological physics [26,27]. The recently discovered type-II hyperbolic lattice, which features both outer and inner edges, provides an ideal platform for addressing this gap [23,36,37]. Although the dynamic transfer of chiral edge states (CESs) in such a system has been theoretically predicted [23], its experimental realization has been impeded by stringent and intricately tailored coupling requirements. Consequently, developing a more experimentally feasible type-II hyperbolic topological model and using it to directly observe the dynamic topological phenomena are highly sought after.

Besides the spatial topology relying on energy gaps and spatial interfaces, recent advancements in photonic time [28-32] and space-time crystals [33-36] have unveiled fascinating temporal topological interface states [28,32,35,36] and space-time topological events [35,36] that originate from the momentum [28,29,32,35,36] and energy-momentum band gaps [35,36], respectively. These fascinating results motivate deeper exploration at the intersection of topological physics and spatiotemporal frameworks. However, time- and space-time topologies have thus far been primarily limited to Euclidean space-time with only one spatial and one temporal dimension ((1+1)-D) [28,30,32,35,36]. In contrast, higher-dimensional hyperbolic spacetime and its accompanying novel spatiotemporal topological phenomena remain completely unexplored.

Here, we present a systematic exploration of the space and space-time topologies in a type-II hyperbolic lattice. By mapping the celebrated Haldane model onto a type-II hyperbolic lattice, we experimentally realize the first type-II hyperbolic Chern insulator and directly observe degenerate outer and inner CESs with opposite chirality. Furthermore, by coupling the counterpropagating outer and inner CESs, we experimentally demonstrate the non-Hermitian phase transition from an anti-parity-time-symmetry-broken (APT-B) phase to an exceptional point (EP), enabling dynamic transfer between the outer and inner CESs with arbitrary proportions. Notably, by engineering the APT phase in a single modulated type-II hyperbolic Chern insulator, we construct a (2+1)-dimensional ((2+1)-D) hyperbolic space-time crystal with an intertwined space Chern and time winding topology. This non-trivial space-time topology gives rise to a unique topological state that is temporally localized at the time interface and spatially localized along the lines on the outer and inner space boundaries, thus forming a novel space-time topological string.

**Results**

**Space topology in a type-II hyperbolic Chern insulator**

We begin with a one-sheet hyperboloid (upper panel of Fig. 1a), which can be continuously projected as a planar hyperbolic model, i.e., "Poincaré ring", preserving the local constant negative curvature and global topology [23,37,38]. This Poincaré ring can be further discretized as a type-II hyperbolic lattice (lower panel of Fig. 1a) via regular polygon tessellation [23,37,38]. The unique geometric degree of freedom of a type-II hyperbolic lattice can be fully characterized by an extended Schläfli symbol $\{\rho_h, p, q\}$, where $p$ and $q$ represent $q$ copies of $p$-sided polygons meeting at each vertex, and $\rho_h$ represents the characteristic radius. We map the celebrated Haldane model onto a type-II hyperbolic $\{0.584, 8, 3\}$ lattice, as illustrated in Fig. 1b. In the light of hyperbolic band theory (HBT), a hyperbolic lattice without spatial translational symmetry can be effectively described in a higher-dimensional momentum space utilizing a hyperbolic Bravais lattice. Following this insight, we select a $\{0.584, 8, 8\}$ Bravais lattice (purple lines) for our type-II hyperbolic Haldane model (see details in Supplementary Note 1). The corresponding unit cell (yellow region) is an octagon containing 16 sites, where the onsite potential of subsite $A$ ($B$) is $+M$ ($-M$), the coupling magnitude between nearest-neighbor (NN) sites is $t_1$, and the coupling magnitude (phase) between next-nearest-neighbor (NNN) sites is $t_2$ ($\phi$). This unit cell is equipped with a Bloch vector $\mathbf{k} = (k_1, k_2, k_3, k_4)$ under a noncommutative translation group $\Gamma = \langle \gamma_1, \gamma_2, \gamma_3, \gamma_4 : \gamma_1 \gamma_2^{-1} \gamma_3 \gamma_4^{-1} \gamma_1^{-1} \gamma_2 \gamma_3^{-1} \gamma_4 \rangle$, whose components form a four-dimensional (4D) Brillouin zone (BZ). We henceforth set the system parameters $t_1 = 1$, $t_2 = 0.2$, $\phi = \pi/2$ and $M = 0$. Figure 1c presents the bulk density of states (DOS) as a function of energy for the unit cell in Fig. 1b, computed using HBT by sampling $\mathbf{k}$ over the 4D BZ. The yellow energy regions with vanished DOS indicate two energy gaps with non-zero Chern vectors $\mathbf{C} = (-1, 1)$ (see details in Supplementary Note 2), revealing a nontrivial Chern topology.

To demonstrate the CESs associated with these Chern gaps, we construct a finite type-II hyperbolic Haldane model with open boundary conditions along the $k_2$, $k_3$ and $k_4$ directions, and periodic boundary conditions along the $k_1$ direction, as illustrated in Fig. 2a. Figure 2b displays the corresponding band structure calculated using HBT, showing the chiral dispersions of the outer (red lines) and inner (green lines) edge modes, spanning the Chern gaps I and II (yellow regions). Figures 2c and 2d respectively show the eigenmode profiles and point-source-excited wave function amplitude distributions of degenerate modes (labeled by "①" and "②" in Fig. 2b) at a mid-gap energy $E_2 = -1.178$ (or $E_1 = 1.178$), verifying a counterclockwise (CCW) or clockwise (CW) unidirectional propagation of outer (upper panel) or inner (lower panel) edge-dominated state.

To experimentally demonstrate the degenerate outer and inner CESs with opposite chirality, we implement the type-II hyperbolic Haldane model to a type-II hyperbolic Chern circuit, as shown in Fig. 2e. Four circuit nodes are connected as a loop via capacitors $C_0$ to form an effective lattice site. Voltages on these four nodes are defined as $V_{i,1}$, $V_{i,2}$, $V_{i,3}$ and $V_{i,4}$, which can be used to construct a pair of pseudospins $V_{\uparrow i,\downarrow i} = V_{i,1} \pm iV_{i,2} - V_{i,3} \mp iV_{i,4}$ to realize the required site couplings. As shown in the inset of Fig. 2e, four pairs of adjacent nodes are directly linked via capacitors $C_1$ to implement the real NN couplings, while four pairs of adjacent nodes are crossly linked via capacitors $C_2$ to implement the complex NNN couplings, and each node is grounded by an inductor $L$. Additionally, all boundary nodes are grounded by extra capacitors to ensure the same resonance frequency as bulk nodes. Based on the Kirchoff's law, we can derive the electrical circuit eigenequation which is consistent with the eigenequation of the type-II hyperbolic Haldane model, identifying the tight-binding model (TBM) parameters in terms of circuit elements as: $t_1 = C_1/C_0$, $t_2 = C_2/C_0$, $\phi = \pi/2$, $E = f_0^2/f^2 - 2 - (3C_1 + 6C_2)/C_0$, where $f$ is the circuit frequency and $f_0 = (2\pi\sqrt{LC_0})^{-1}$ (see details in Supplementary Note 3). Specifically, the electrical circuit element parameters are selected as $C_0 = 1.5\text{nF}$, $C_1 = 1.5\text{nF}$, $C_2 = 0.3\text{nF}$, and $L = 10\mu\text{H}$.

We first experimentally measure the impedance spectra at a bulk node, an outer edge node, and an inner edge node (illustrated by grey, red and green dots) in type-II hyperbolic Haldane circuit, as shown in Fig. 2f. We identify two gaps (yellow regions) in frequency ranges of 0.45-0.496 MHz (gap I) and 0.554-0.62 MHz (gap II), within which the outer and inner edge nodes exhibit significant impedance responses while the bulk node shows negligible impedance responses. Additionally, we measure the spatial impedance distributions of the type-II hyperbolic Haldane circuit at frequency $f_1 = 0.47$ MHz in gap I and $f_2 = 0.586$ MHz in gap II, respectively, as shown in Fig. 2g and 2h. These edge-dominated profiles match well with the superposition of the two edge eigenmodes in Fig. 2c, confirming the coexistence of outer and inner edge states. Next, we apply a chiral voltage source (cyan star) $[V_{i,1}, V_{i,2}, V_{i,3}, V_{i,4}] = [V_0, iV_0, -V_0, -iV_0]\sin(2\pi f_s t)$ ($V_0 = 0.5V$) at an outer or inner edge node to selectively excite the spin-up voltage pseudospin to demonstrate the unidirectionality of these hyperbolic edge states. Figure 2i shows the measured voltage amplitude distributions of the outer (upper panel) and inner (lower panel) edge states at frequency $f_2$ in gap II. It can be seen that the outer (inner) edge voltage field is tightly confined to the outer (inner) edge and propagates unidirectionally in a CCW (CW) direction. Additionally, the robustness of these CESs against defects has also been experimentally examined (see details in Supplementary Note 4).

**Anti-parity-time phase transition of coupled counterpropagating CESs**
Next, we explore the non-Hermitian APT phase transition by coupling the counterpropagating outer and inner CESs in a modulated type-II hyperbolic Chern insulator. We start with a two-level model and employ the outer and inner CESs in Fig. 2c as its orthonormal basis $\{|\psi_o\rangle, |\psi_i\rangle\}$, whose purely CCW or CW power flows are defined as $\beta_o = |\langle\psi_o|\psi_o\rangle|^2 - |\langle\psi_i|\psi_o\rangle|^2 = 1$ and $\beta_i = |\langle\psi_o|\psi_i\rangle|^2 - |\langle\psi_i|\psi_i\rangle|^2 = -1$, respectively. To manipulate the couplings between these bases, we introduce a radial coupling channel (blue region) consisting of four modulated couplings (short red lines) $mt_1$ into type-II hyperbolic Haldane model, as illustrated in Fig. 3a. We adopt an outer or inner edge source to excite the coupled CESs $|\psi_{m,k}\rangle$ ($k = 1, 2$) (see Supplementary Note 5) and define the corresponding power flow as $\beta_k = |\langle \psi_o|\psi_{m,k}\rangle|^2 - |\langle \psi_i|\psi_{m,k}\rangle|^2$. Figure 3b presents the $\beta_k$ as a function of the modulation strength $m$, showcasing that as $m$ increases, $\beta_k$ simultaneously changes from 1 (red circles) or -1 (blue circles) to 0. Additionally, $\beta_k$ can also be retrieved as a function of effective coupling strength $\kappa$, as shown by circles in Fig. 3c. This phenomenon of power flow evolution can be effectively modeled by a Hamiltonian $H_{APT} = \begin{pmatrix} \beta_o & i\kappa \\ i\kappa & \beta_i \end{pmatrix} = \begin{pmatrix} 1 & i\kappa \\ i\kappa & -1 \end{pmatrix}$. Note that its apparent non-Hermiticity of the purely imaginary coupling does not originate from any physical gain or loss. Instead, it is an pseudo-Hermiticity, i.e., $\sigma_z H_{APT} \sigma_z^{-1} = H_{APT}^\dagger$ with $\sigma_z = \begin{pmatrix} 1 & 0 \\ 0 & -1 \end{pmatrix}$, which follows from power-flow-difference conservation, a unique form of energy-conservation inherent to contra-directional coupling [23-25] (see Supplementary Note 6). This Hamiltonian is also anti-parity-time symmetric (APT-S), i.e., $\{H_{APT}, \hat{P}\hat{T}\} = 0$, and gives eigenvalues:

$$\beta = \pm\sqrt{1 - \kappa^2} \tag{1}$$

The analytic solutions are plotted as solid lines in Fig. 3c, exhibiting a good quantitative agreement with the numerical results (red and blue circles). Specifically, in the weakly coupled regime $|\kappa| < 1$ (blue region), $\beta$ appear as two real numbers (purple dashed line), supporting two oppositely propagating eigenstates that correspond to the APT-B phase. The defining feature of this phase is that an excitation applied to either edge couples only partially to the opposite edge through the coupling channel. As a result, the wave function amplitude remains predominantly localized on the initial edge, as shown by the two modes (labeled "①" and "②") with $m = 6$ in the left insets of Fig. 3b. When the coupling reaches a critical value $|\kappa| = 1$ (red dashed line), $\beta$ coalesces to zero, indicating the formation of two coalesced non-propagating eigenstates that correspond to the EP (red star). The hallmark of this point is that an excitation on either edge couples completely to the opposite edge via the coupling channel. In other words, the two edges become fully connected regardless of the excitation position, leading to a wave function amplitude equally distributed between both edges, as shown by the two modes (labeled "③" and "④") with $m = 41$ in the right insets of Fig. 3b. Figure 3d displays the local DOS distributions of the modes for $m = 6$ (upper panel) and $m = 41$ (lower panel), which correspond to the superposition of two APT-B modes and two EP modes in Fig. 3b, respectively. The key difference between the two mode profiles is that, in the former case, the amplitude at the coupling channel is significantly weaker than that at the edges, whereas in the latter case, the amplitude at the coupling channel is comparable to that at the edges, indicating that the two edges are partially and fully coupled, respectively. Note that these features persist even in the presence of losses, although attenuation modifies the wave-function amplitude distributions. Figures 3e and 3f show the wave function amplitude distributions with propagation loss $\eta = 0.04$, excited by an outer or inner edge source (cyan star) with energy $E_2$ for $m = 6$ and $m = 41$, respectively. It can be seen that, in the APT-B phase ($m = 6$), a CCW (CW) CES at the outer (inner) edge couples only partially to the CW (CCW) CES at the inner (outer) edge via the coupling channel; the wave function amplitude remains predominantly localized at the outer (inner) edge despite loss-induced attenuation. By contrast, at the EP ($m = 41$), a CCW (CW) CES at the outer (inner) edge couples completely back to a CW (CCW) CES at the inner (outer) edge via the coupling channel; however, the wave-function amplitude is no longer equally distributed between the two edges owing to loss-induced attenuation.

To experimentally demonstrate the APT phase transition of the coupled outer and inner CESs, we construct a radial coupling channel by parallelling five additional capacitors $C_{m_1} = C_0$ ($C_{m_2} = 8C_0$) to each of the four NN couplings to enhance their effective TBM parameter as $t_{m_1} = (5C_{m_1} + C_1)/C_0 = 6$ ($t_{m_2} = (5C_{m_2} + C_1)/C_0 = 41$), as illustrated in the inset of Fig. 3g, corresponding to a modulation strength $m = 6$ ($m = 41$). Figures 3h and 3i display the measured spatial impedance distributions of the modulated type-II hyperbolic Haldane circuit, i.e., effective local DOS, at frequency $f_2$ for $m = 6$ and $m = 41$, respectively, whose mode profiles match well with the local DOS profiles shown in Fig. 3d. Figures 3j and 3k shows the measured voltage amplitude distributions of the coupled CESs excited by an outer or inner edge source (cyan star) at frequency $f_2$ for $m = 6$ and $m = 41$, respectively. For the weak manipulation ($m = 6$), the outer (inner) CES partially transfers into the inner (outer) CES through the coupling channel, corresponding to an APT-B phase. While for the strong manipulation ($m = 41$), the outer (inner) CES completely turn into the inner (outer) CES through the coupling channel, corresponding to an EP, thereby achieving the dynamic transfer of the outer and inner CESs.

**Hyperbolic space-time topological string**
Finally, by precisely engineering the APT phase governing pulse dynamics in a single double-channel-modulated type-II hyperbolic Chern insulator, we construct a discrete (2+1)-D hyperbolic space-time crystal; with additional non-Hermitian modulations, we further demonstrate a space-time topological string. As illustrated in Fig. 4a, we introduce two distinct radial coupling channels $C_1$ and $C_2$ and two edge modulators $M_S$ and $M_L$ into a {0.9984, 8, 3} type-II hyperbolic Haldane model. Specifically, the modulation strengths of $C_1$ and $C_2$ are set to $m = 8.3805$ and $m = 3001$, respectively, and $M_S$ and $M_L$ consist of edge on-site gain/loss and a phase shifter. For a Gaussian pulse with a specific width and an energy centered at $E_2 = -1.178$ (see details in Supplementary Note 7), $C_1$ and $C_2$ effectively function as 50:50 and 0:100 beam splitters, respectively, with an output phase difference of $\pi/2$ (see details in Supplementary Note 8). Meanwhile, $M_S$ and $M_L$ effectively operate as pulse modulators that provide a composite modulation $e^\varphi = e^{\varphi_a + i\varphi_p}$, where $\varphi_a$ and $\varphi_p$ denote the amplitude and phase modulation strengths (see details in Supplementary Notes 9 and 10), as illustrated in Fig. 4b. In this way, the outer and inner edges are partitioned into two closed loops with distinct path lengths, denoted as loop-S (the shorter loop, marked in purple) and loop-L (the longer loop, marked in red). The two closed loops are separated at $C_2$ and connected through $C_1$, and are individually modulated by $M_S$ and $M_L$, respectively. By defining a loop unit length $\Delta$, the lengths of loop-S and loop-L are set as $L_S = 19.5\Delta$ and $L_L = 20.5\Delta$, respectively (see details in Supplementary Note 11). Interestingly, to simplify the analysis without considering non-Hermitian modulations, when a single pulse is injected into loop-L, it undergoes iterative splitting and recombination at $C_1$, consequently evolving into two pulse trains (one in each loop). Figure 4c shows the snapshots of intensity distribution of pulse trains in the edge loops at the first four time-steps $\mathsf{t} = 0T, 1T, 2T, 3T$, where $T = (L_S + L_L)/2$ represents the average time period for a pulse with a unit velocity propagating along the loop-S and loop-L. The corresponding intensity functions of sub-pulses in the loop-L (loop-S) with respect to the loop length $s \in [0, s_0]$ (i.e., pulse counting region), $|\mathbb{u}_\mathsf{t}(s)|^2$ ($|\mathbb{v}_\mathsf{t}(s)|^2$), are displayed in the upper (lower) panel of each row in Fig. 4d. Owing to the length difference between two loops, within a time interval of $\mathsf{t} = t \times T$, the evolved sub-pulses in loop-S or loop-L (black dashed lines) acquire negative or positive positional shifts with respect to the initial pulse (red dashed lines), with the displacement quantified in units of $\Delta s = \Delta/2$. Remarkably, these shifts essentially index the relative positions of each pulse along the pulse trains, which map to the discrete space coordinate $x$, and the count of time-step of the pulse train maps to the discrete time coordinate $t$. By extracting the local maxima of $|\mathbb{u}_\mathsf{t}(s)|^2$ ($|\mathbb{v}_\mathsf{t}(s)|^2$) and inserting zeros between adjacent maxima, $|\mathbb{u}_\mathsf{t}(s)|^2$ ($|\mathbb{v}_\mathsf{t}(s)|^2$) can be discretely reduced into $|u_x^t|^2$ ($|v_x^t|^2$) ($t = 0, 1, 2, 3$), as displayed in the upper (lower) panel of each row in Fig. 4e. The corresponding spatiotemporal distributions of the total pulse intensity, defined as $P(x, t) = |u_x^t|^2 + |v_x^t|^2$, are demonstrated in Fig. 4f. Particularly, another equivalent coordinate description regarding time delay of this form of time-multiplexing is that: after $t$ round-trips evolutions, the sub-pulses in loop-S or loop-L accumulate the total time delay $t \times T \pm x \times \Delta t = t \times T \pm x \times \Delta s$, where the positive and negative signs indicate the relative time delay and time advance, respectively, thereby yielding a (1+1)-D synthetic space-time lattice consist of a mesh of beam splitters (BS (50:50)), as illustrated in Fig. 4g. Together with the discrete radial spatial distance $\rho$, the pulse trains effectively evolve in a discrete (2+1)-D hyperbolic spacetime $(\rho, x, t)$. Notably, the space-time topology in the spacetime $(\rho, x, t)$ is essentially the intertwining of a space topology defined in the $(\rho, x)$ sub-spacetime, ensuring the formation of spatial topological edge states, and a time topology defined in the $(x, t)$ sub-spacetime, determining the emergence of temporal topological interface states.

As illustrated in Fig. 5a, the pulse dynamics in a synthetic lattice with non-Hermitian modulations is governed by the recursive evolution equations (see details in Supplementary Note 11):

$$u_x^{t+1} = \frac{1}{\sqrt{2}}(-u_{x+1}^t + iv_{x+1}^t)e^{\varphi_u(x,t+1)}$$

$$v_{x+2}^{t+1} = \frac{1}{\sqrt{2}}(iu_{x+1}^t - v_{x+1}^t)e^{\varphi_v(x+2,t+1)} \tag{2}$$

The modulation parameter $\varphi_u(x,t)$ ($\varphi_v(x,t)$) setting for $u_x^t$ ($v_x^t$) is set as amplitude modulation only, i.e., $\varphi_u = \varphi_{u_a}$ ($\varphi_v = \varphi_{v_a}$). We select a unit cell (green shaded region) that has a two-step periodicity in space, accompanied by a four-step periodicity in time with gain-loss modulation settings $[\varphi_u(x,t), \varphi_u(x+1,t+1), \varphi_u(x,t+2), \varphi_u(x+1,t+3)] = [g, 0, 0, -g]$ and $\varphi_v = -\varphi_u$, where $g$ is the gain/loss strength. For the Hermitian case, i.e., $g = 0$, by applying Floquet-Bloch ansatz to Eq. (2), we can obtain the system's band structure, as shown in Fig. 5b, which is gapless in both energy and momentum. While for the non-Hermitian case, e.g., $g = 0.3$ ( $g = -0.3$ ), two momentum bandgaps (purple and blue regions) open at momenta $k = 0$ and $\pi$ (see details in Supplementary Note 12), as shown in Fig. 5h (Fig. 5i). The topology of the momentum bandgap can be characterized by a temporal winding number:

$$\nu_t = \frac{i}{2\pi} \sum_{s=1}^{2} \int_{-\pi}^{\pi} \left\langle \psi_{k_s}(k) \left| \frac{\partial}{\partial E} \right| \psi_{k_s}(k) \right\rangle dE \tag{3}$$

where $\psi_{k_s}(k)$ represents the eigenstates of two momentum bands. Specifically, for $g = 0.3$, $\nu_t = +1$, while for $g = -0.3$, $\nu_t = -1$ (see details in Supplementary Note 13).

For the entire (2+1)-D hyperbolic space-time, in the Hermitian case ($g = 0$), without the time winding topology, the space Chern topology supports a space-time topological world sheet, where topological states spatially localized on the outer and inner boundaries ($x$-$t$ surfaces) and remain temporally delocalized. To demonstrate this spatiotemporal topological state, we construct a finite discrete (2+1)-D hyperbolic space-time which possesses a pure space Chern topology, as illustrated in Fig. 5c. The corresponding values of $g$ over time are shown in Fig. 5d. We initially inject a spatially broad wave-packet with a momentum $k = 0$ from the loop-L at time $\mathfrak{t} = 0$. The evolution of the intensity function of the sub-wave-packet in the loop-L (loop-S) $|\mathbb{u}_\mathfrak{t}(s)|^2$ ($|\mathbb{v}_\mathfrak{t}(s)|^2$) over time is displayed in the left (right) panel of Fig. 5e, from which we can see that the topological states are spatially localized on the outer and inner boundaries. Figure 5f shows the spatiotemporal distribution of the total wave-packet intensity $P(x,t)$. It can be seen that the total wave-packet intensity $P(t) = \sum_x(|u_x^t|^2 + |v_x^t|^2)$ keep almost unchanged over time, as shown in Fig. 5g, indicating the topological states remain temporally delocalized. In the non-Hermitian case ($g \neq 0$), the time winding topology is intertwined with the space Chern topology, the resulting space-time topology of this (2+1)-D hyperbolic space-time is characterized by a hybrid topological invariant $(C, \nu_t)$. This invariant guarantees the emergence of a space-time topological string, i.e., a time-constant slice of the space-time topological world sheet, where a topological state is not only spatially localized on the outer and inner boundaries but also temporally localized at a specific time $t$ slice. To demonstrate this space-time topological string state, we construct a finite discrete (2+1)-D hyperbolic space-time crystal which consists of regions I and II possessing space-time topologies of $(C, 1)$ and $(C, -1)$, respectively, thereby forming a temporal interface, as illustrated in Fig. 5j. The corresponding values of $g$ over time are shown in Fig. 5k, where the black dashed line represents the temporal interface. Figures 5l and 4m show $|\mathbb{u}_t(s)|^2$ and $|\mathbb{v}_t(s)|^2$ over time and $P(x,t)$ across all $(x,t)$ coordinates, respectively. It can be seen that the wave-packet is localized at the time-topological interface where $\nu_s$ flips its sign, and $P(t)$ exhibits exponential growth and decay before and after the temporal interface, as shown in Fig. 5n. This spatiotemporal topological state is exponentially localized along both the spatial axis $\rho$ and temporal axis $t$, and extended along spatial axis $x$, signifying the forming of a space-time topological string. Additionally, the robustness of this space-time topological string against temporal disorders has also been examined (see details in Supplementary Note 14).

**Discussion**
In conclusion, we have unveiled space and space-time topologies in a type-II hyperbolic lattice. For the space topology, we extended the celebrated Haldane model to a type-II hyperbolic lattice and experimentally realized a type-II hyperbolic Chern insulator with degenerate counterpropagating CESs at its outer and inner edges. Furthermore, we experimentally demonstrated the APT phase transition by coupling the outer and inner CESs of the type-II hyperbolic Chern insulator, achieving dynamic transfer of these CESs with arbitrary proportions. For the space-time topology, by engineering the APT phase in a single double-channel-modulated type-II hyperbolic Chern insulator, we proposed a novel scheme to construct a (2+1)-D hyperbolic space-time crystal with combined space Chern and time winding topologies, resulting in a unique space-time topological string state. Our work not only expands the horizon of hyperbolic topological physics but also establishes a framework for spatiotemporal manipulation of hyperbolic topological states. These advances pave the way for robust, compact, and efficient hyperbolic topological devices, such as hyperbolic topological lasers [39,40] and optical frequency combs [41-43].

**Data availability**
All data that support the plots within this paper and other findings of this work are available from the corresponding authors upon reasonable request.

**Code availability**
All the codes used to generate and/or analyze the data in this work are available from the corresponding authors upon reasonable request.

**Acknowledgments**
Z.G. acknowledges fundings from the National Key R&D Program of China (grant no. 112025YFA1412300), National Natural Science Foundation of China (grant no. 62361166627 and 62375118), Guangdong Basic and Applied Basic Research Foundation (grant no. 2024A1515012770), Shenzhen Science and Technology Innovation Commission (grants no. 20230802205352003), and High-level Special Funds (grant no. G03034K004). Z.B.Z. acknowledges the funding from the China Postdoctoral Science Foundation (grant No. 2025M773439). Z. Z. acknowledges the funding from the China Postdoctoral Science Foundation (grant No. 2025M773439).

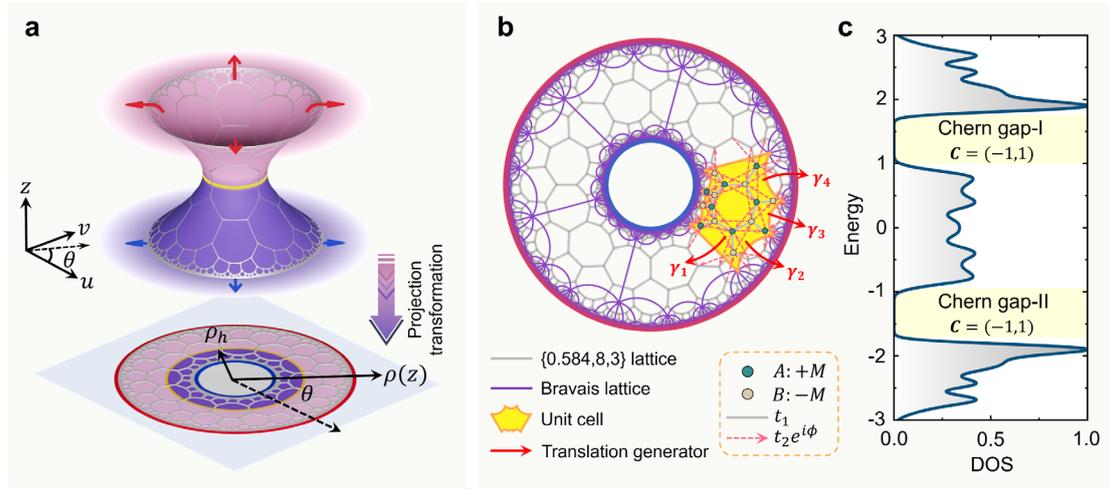

**Fig. 1 | Type-II hyperbolic Chern insulator. a** A periodic lattice on a one-sheet hyperboloid ($u^2 + v^2 - z^2 = 1$) in (1+2)-dimensional ($z, u, v$) Minkowski spacetime can be mapped onto a type-II hyperbolic lattice on a Poincaré ring via an isometric projective transformation. **b** Schematic illustration of the Haldane model on an infinite type-II hyperbolic {0.584, 8, 3} lattice. The grey lines represent the type-II hyperbolic {0.584, 8, 3} lattice, and the purple lines represent the corresponding {0.584, 8, 8} Bravais lattice. The unit cell is highlighted in yellow, which consists of 16 sites connected via real nearest-neighbor couplings (grey lines) and complex next-nearest-neighbor couplings (pink dashed arrows). The red arrows labeled $\gamma_1, \gamma_2, \gamma_3$ and $\gamma_4$ are translation generators of the Bravais lattice. **c** Bulk density of states for the unit cell in **b**, which are computed using the hyperbolic band theory with $M = 0, t_1 = 1, t_2 = 0.2$ and $\phi = \pi/2$, revealing two Chern gaps (light yellow regions) with non-zero Chern vectors $C$.

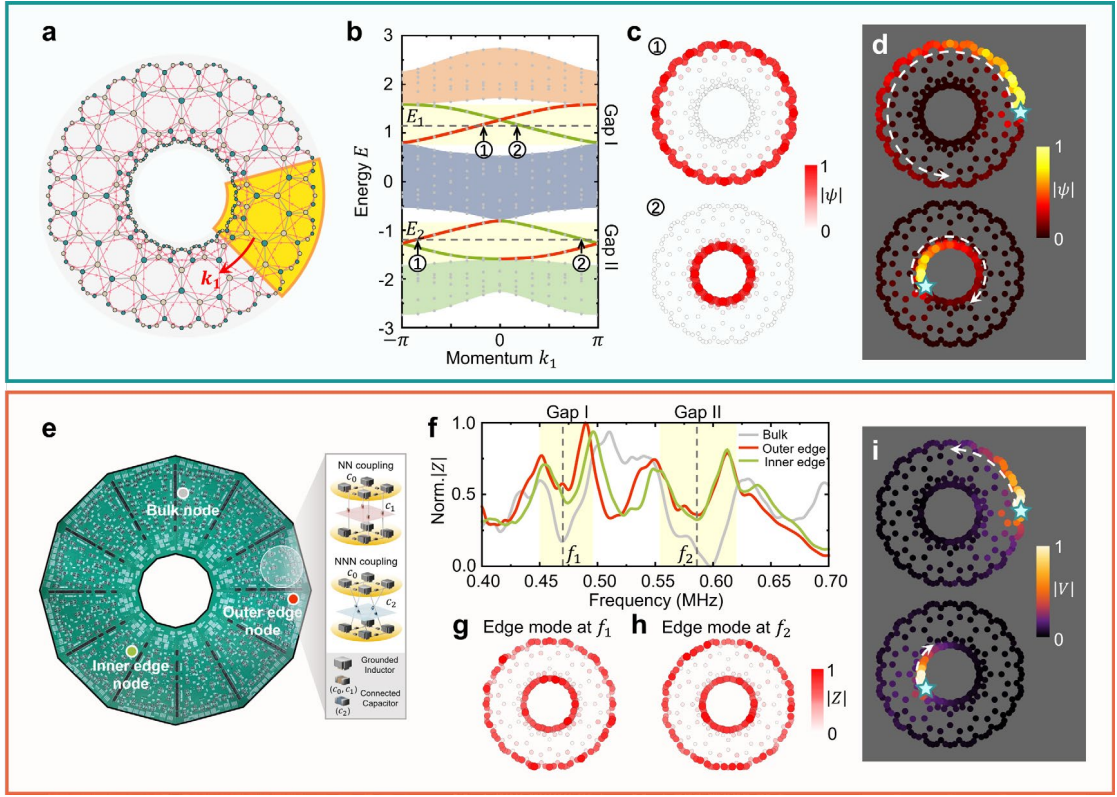

**Fig. 2 | Space topology in a type-II hyperbolic Chern insulator. a** Schematic illustration of a finite type-II hyperbolic {0.584, 8, 3} Haldane model with 288 sites in total. The yellow region represents a unit cell. **b** Band structure of the type-II hyperbolic {0.584, 8, 3} Haldane model in **a**, which is computed using the hyperbolic band theory. The red (green) lines in gaps I and II (light yellow regions) represent the dispersion of the outer (inner) CESs. **c** Mode profiles of outer (upper panel) and inner (lower panel) edge states labeled "①" and "②" in **b**. **d** Calculated wave function amplitude distributions of outer (upper panel) and inner (lower panel) CESs excited by an edge source (cyan star) with energy $E_2 = -1.178$. A small loss $\eta = 0.04$ is introduced to distinguish the chirality of CESs. **e** Photograph of the fabricated type-II hyperbolic Haldane circuit. The inset illustrates the effective lattice site (yellow disks), real nearest-neighbor coupling, and imaginary next-nearest-neighbor coupling. **f** Measured impedance spectra at a bulk node (grey line), an outer edge node (red line), and an inner edge node (green line), illustrated by colored dots in **e**. The yellow regions correspond to the two gaps in **b**. **g, h** Measured impedance distributions at frequency $f_1 = 0.47$MHz in gap I (**g**) and $f_2 = 0.586$MHz in gap II (**h**), respectively. **i** Measured voltage amplitude distributions of outer (upper panel) and inner (lower panel) CESs excited by an edge source (cyan stars) at frequency $f_2$.

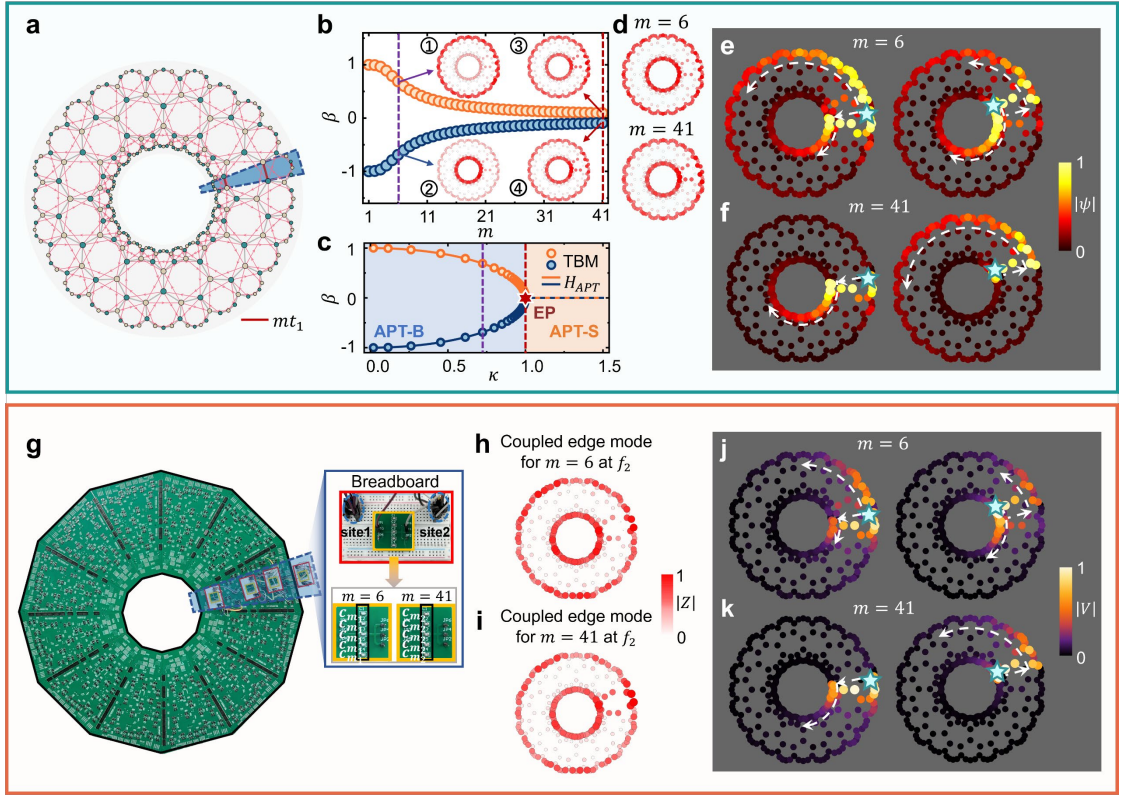

**Fig. 3 | Anti-parity-time (APT) phase transition and dynamic transfer of CESs in a single-channel-modulated type-II hyperbolic Chern insulator. a** Schematic illustration of a single-channel-modulated type-II hyperbolic Haldane model. The blue region represents a radial coupling channel consisting of four modulated nearest-neighbor couplings $mt_1$ (red short lines). **b** The power flow $\beta$ as a function of modulation strength $m$. The insets display the mode profiles of coupled CESs for $m = 6$ (labeled "①" and "②") and $m = 41$ (labeled "③" and "④"), respectively. **c** The power flow $\beta$ as a function of effective coupling factor $\kappa$. The solid lines (circles) are analytic (numerical) results calculated from the Hamiltonian $H_{APT}$ (tight-binding model). The blue and orange regions denote the anti-parity-time symmetry-broken (APT-B) and anti-parity-time symmetric (APT-S) phases, respectively. The red star represents an exceptional point (EP). **d** Local DOS of the superimposed modes for $m = 6$ (upper panel) and $m = 41$ (lower panel), respectively. **e, f** Calculated wave function amplitude distributions of coupled CESs excited by an outer (left panel) or inner (right panel) edge sources (cyan star) with energy $E_2 = -1.178$ for $m = 6$ (**e**) and $m = 41$ (**f**), respectively. **g** Photograph of the modulated type-II hyperbolic Haldane circuit. The inset depicts the experimental implementation of the radial coupling channel for $m = 6$ and $m = 41$, respectively. **h, i** Measured impedance distributions at frequency $f_2 = 0.586 \text{MHz}$ for $m = 6$ (**h**) and $m = 41$ (**i**), respectively. **j, k** Measured voltage amplitude distributions of coupled CESs excited by an outer (left panel) or inner (right panel) edge source (cyan stars) at frequency $f_2$ for $m = 6$ (**j**) and $m = 41$ (**k**), respectively.

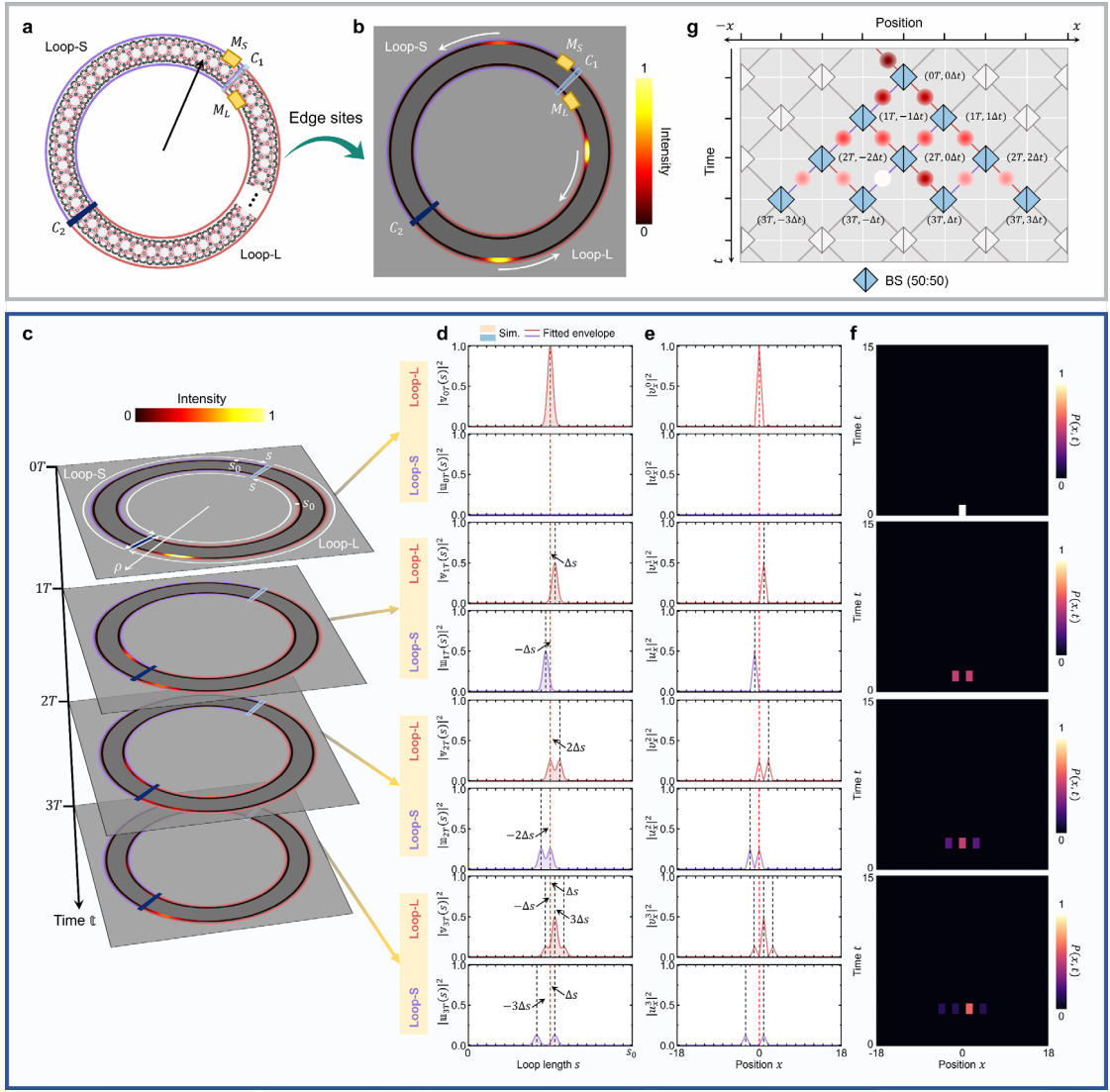

**Fig. 4 | Synthetic space-time lattice mapped from the pulse evolution in a double-channel-modulated type-II hyperbolic Chern insulator. a** Schematic illustration of a double-channel-modulated type-II hyperbolic {0.9984, 8, 3} Haldane model. The light (dark) blue region represents the radial coupling channel $C_1$ ($C_2$). The orange regions represent the edge modulators $M_1$ and $M_2$. The purple (red) arcs outline the loop-S (loop-L). **b** Schematic illustration of the splitting and amplification/attenuation of a pulse in the edge loops in **a**. **c** Snapshots of intensity distribution of the evolved pulse trains at times $\mathbb{t} = 0T, 1T, 2T, 3T$. The loop-S and loop-L are effectively parameterized by the loop length $s \in [0, s_0]$. **d** Intensity functions of sub-pulses $|\mathbb{v}_\mathbb{t}(s)|^2$ (upper panel) and $|\mathbb{u}_\mathbb{t}(s)|^2$ (lower panel) in the loop-L and loop-S at each time step in **c**. The red and black dashed lines label the relative positions of the initial and evolved pulses, respectively. **e** Discrete intensity functions of sub-pulses $|v_x^t|^2$ (upper panel) and $|u_x^t|^2$ (lower panel) in the loop-L and loop-S at each time step in **c**. **f** Spatiotemporal distribution of the total pulse intensity $P(x,t)$. **g** (1+1)-D synthetic space-time lattice which consists of a mesh of BS (50:50).

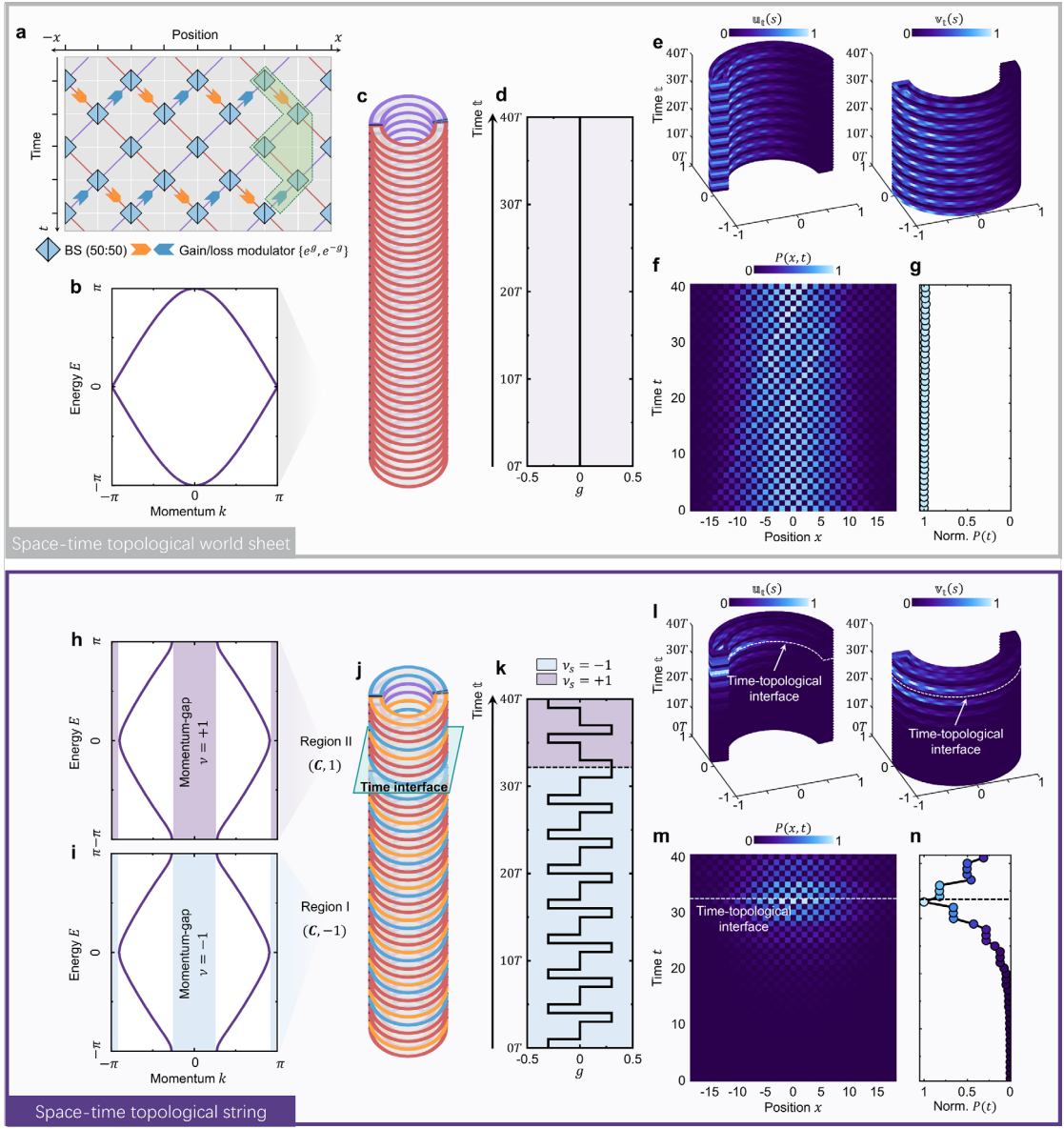

**Fig. 5 | Hyperbolic space-time topological string. a** (1+1)-D synthetic space-time lattice which consists of a mesh of BS (50:50) and gain-loss modulations ($e^{\pm g}$). The green shaded region represents a unit cell with a four-time-step and a two-space-step period. **b-d** Discrete (2+1)-D hyperbolic space-time with a pure space Chern topology (**c**), where the temporal setting of $g$ is shown in **d**, corresponding to the synthetic lattice in **a** with $g = 0$ whose band structure is shown in **b**. **e** Evolutions of the intensity functions of sub-wave-packet $|\mathbb{u}_\mathbb{t}(s)|^2$ and $|\mathbb{v}_\mathbb{t}(s)|^2$ in the loop-S (left panel) and loop-L (right panel) of **c** over time, respectively. **f** Spatiotemporal distribution of the total pulse intensity $P(x,t)$ extracted from **e**. **g** Total pulse intensity $P(t)$ of **f** as a function of time. **h-k** Discrete (2+1)-D hyperbolic space-time composing regions I and II with space-time topologies of $(C,-1)$ and $(C,1)$, respectively, forming a temporal interface (**j**), where the temporal setting of $g$ is depicted in **k**. Region I (II) corresponds to the synthetic lattice in **a** with $g = -0.3$ ($g = 0.3$), whose band structure is shown in **h** (**i**). **l** Evolutions of the intensity functions of sub-wave-packet $|\mathbb{u}_\mathbb{t}(s)|^2$ and $|\mathbb{v}_\mathbb{t}(s)|^2$ in the loop-S (left panel) and loop-L (right panel) of **j** over time, respectively. **m**

Spatiotemporal distribution of the total pulse intensity $P(x, t)$ extracted from **l**. **n** Total pulse intensity $P(t)$ of **m** as a function of time.